\documentclass[aps,reprint,amsmath,amssymb,nofootinbib]{revtex4-1}
\usepackage{graphicx}
\usepackage{slashed}
\usepackage{float}
\usepackage{hyperref}
\usepackage{bm}
\usepackage{url}
\usepackage{accents}
\usepackage{scrextend}

\DeclareMathOperator{\e}{e}
\def\p{\bm{p}}
\def\dual#1{\accentset{\boldsymbol{\neg}\vspace{-.13ex}}{#1}}

\begin{document}

\bibliographystyle{plain}

\title{ELKO fermions as dark matter candidates}

\author{Bakul Agarwal$^{1,}$}
\email{agarwalb@msu.edu}
\author{Pankaj Jain$^{2,}$}
\email{pkjain@iitk.ac.in}
\author{Subhadip Mitra$^{2,3,}$}
\email{subhadipmitra@gmail.com}
\email{subhadip.mitra@iiit.ac.in}
\author{Alekha C. Nayak$^{2,}$}
\email{acnayak@iitk.ac.in}
\author{Ravindra K. Verma$^{2,}$}
\email{ravindkv@iitk.ac.in}
\affiliation{$^1$ Department of Physics and Astronomy, Michigan State University,
East Lansing, MI, USA 48824} 
\affiliation{$^2$ Department of Physics, Indian Institute of Technology, Kanpur - 208016, India}
\affiliation{$^3$ Center for Computational Natural Sciences and Bioinformatics, International Institute of Information Technology, Hyderabad 500 032, India}

\begin{abstract}
We study the implications of the ELKO fermions as a cold dark matter candidate. Such fermions 
arise in theories that are not symmetric under the full Lorentz group. Although they do not carry electric charge, ELKOs can still couple to photons through a nonstandard interaction. They also couple to the Higgs but do not couple to other standard model particles. We impose limits on their coupling strength and the ELKO mass 
assuming that these particles give dominant contribution to the cosmological cold dark matter. We also determine limits imposed by the direct dark matter 
search experiments on the ELKO-photon and the ELKO-Higgs coupling. 
Furthermore we determine the limit imposed by the gamma ray bursts time
delay observations on the ELKO-Higgs coupling. We find that astrophysical
and cosmological considerations rule out the possibility that ELKO may
contribute significantly as a cold dark matter candidate. The only allowed
scenario in which it can contribute significantly 
as a dark matter candidate is that it
was never in equilibrium with the cosmic plasma. We also obtain a 
relationship between the ELKO self-coupling and its mass by demanding it to be 
consistent with observations of dense cores in the galactic centers. 
\end{abstract}

\maketitle
\section{Introduction}
\label{sec:intro}
Current cosmological observations indicate that cold dark matter (CDM) 
contributes 23\% of the energy density of the Universe. The nature of this 
matter
is so far unknown but  
there are many proposals for dark matter \cite{Bertone:2004pz,Jungman:1995df,Bergstrom:2000pn,Feng:2010gw,G.Bertone}.
In 2005, Ahluwalia and Grumiller proposed a spin half fermion with mass 
dimension one \cite{Ahluwalia:2004ab,Ahluwalia:2004sz}. The field, called 
ELKO, is an 
eigenspinor of the charge conjugation 
operator and and hence carries no electric charge. 
Moreover, the mismatch between the mass dimension of ELKO and the 
standard model (SM) fermions 
restricts its interactions with the SM particles  \cite{Ahluwalia:2010zn} 
making ELKO a suitable candidate for dark matter.

ELKO arises in theories that are not symmetric under the full Lorentz
group \cite{Ahluwalia:2010zn,Gillard:2010nr} but only a subgroup, 
such as SIM(2) \cite{Cohen:2006ky}.
Had the SM respected either P, T, CP or CT, then
the subgroup SIM(2) would necessarily be enhanced to the full Lorentz group
but it breaks these discrete symmetries and 
allows the possibility of a small violation of the Lorentz invariance.    
As Cohen and Glashow \cite{Cohen:2006ky} argued in 2006, 
``Many empirical successes of of special relativity need not demand Lorentz invariance of the underlying framework." 
These theories have a preferred axis 
\cite{Ahluwalia:2010zn,Ahluwalia:2008xi,Cohen:2006ky}, 
that breaks Lorentz invariance by breaking rotational symmetry.
Along such a preferred axis, the ELKO field enjoys locality \cite{Ahluwalia:2009rh}.
It is intriguing that
cosmological observations also show some evidence for a  
preferred axis in the Universe \cite{Ralston:2003pf,Ade:2013nlj}. 

ELKO interacts dominantly with the Higgs field and thus acts as a dark matter 
candidate somewhat analogous to the Higgs portal models, see for example
\cite{Wilczek:2006,Kim:2007,Hooper:2008}. It also has
a quadratic 
self-coupling
as well as 
a coupling to the electromagnetic field tensor $F^{\mu\nu}$.
We find that the electromagnetic coupling is severely restricted by direct dark 
matter searches. 
At the Large Hadron Collider (LHC) its discovery prospects through its Higgs 
interaction \cite{Dias:2010aa,Alves:2014kta,Alves:2014qua} as well as
 possible indirect detection \cite{Gillard:2011mv} have been studied. 
The ELKO spinor 
driven inflation \cite{Boehmer:2006qq,Boehmer:2007dh,Boehmer:2008rz,Basak:2012sn,Boehmer:2010ma,Shankaranarayanan:2009sz,Basak:2011wp,Fabbri:2010ws} and its application to gravity \citep{Fabbri:2010va,Fabbri:2011mi} and higher dimensional brane world model \cite{Liu:2011nb,Jardim:2014cya,Jardim:2014xla} have been 
proposed. The causality \cite{Fabbri:2009aj} structure as well as a dynamical
mass generation mechanism of the 
 ELKO field \citep{Bernardini:2012sc,daRocha:2011yr} has been discussed 
in the literature. The 
ELKO spinor has been shown as a building block of Duffin-Kemmer-Petiau algebra in
Ref. \cite{Cavalcanti:2014uta}.

In the present paper we 
 systematically investigate its implications as a dark matter candidate.
In particular we determine the range of parameters over which it can
act as a CDM candidate. Furthermore we investigate whether
this range is consistent with the known limits on dark matter couplings.
This issue has not been addressed so far in the literature.

This paper is organized as follows: In Sec.~\ref{sec:review}, we briefly review the ELKO field and its interactions. 
In Sec.~\ref{sec:dark}
we determine the range of values of the ELKO mass and its coupling with 
the Higgs for which it may be considered as a CDM candidate.
 In Sec. ~\ref{sec:scatt} we determine the constraints on the
 ELKO-photon and the ELKO-Higgs couplings arising from 
the CDMS II limit on the scattering of dark matter with protons. 
In Sec. \ref{sec:GRB} we obtain the constraints on the ELKO-Higgs
coupling arising from gamma ray bursts. In Sec. 
\ref{sec:self-interaction} we determine the implications of the 
galactic dark matter cores for the ELKO self-coupling. 
Finally we conclude in Sec.~\ref{sec:conclusions}.

\section{A Brief review of elko fermion and its interactions}
\label{sec:review}
The Fourier decomposition of the ELKO field may be written as \cite{Ahluwalia:2013uxa}
\begin{subequations}
\begin{eqnarray}
\mathfrak{f}(x) =  && \int \frac{\text{d}^3p}{(2\pi)^3}  \frac{1}{\sqrt{2 m E(\p)}} \sum_\alpha \Big[ a_\alpha(\p)\lambda^S_\alpha (\p) \exp(- i p_\mu x^\mu)\nonumber \\
&& \hspace{11pt}+\, b^\dagger_\alpha(\p)\lambda^A_\alpha(\p) \exp(i p_\mu x^\mu){\Big]}
\label{eq:elko}
\end{eqnarray}
and its dual as
\begin{eqnarray}
\dual{\mathfrak{f}}(x)  =  && \int \frac{\text{d}^3p}{(2\pi)^3}  \frac{1}{\sqrt{2 m E(\p)}} \sum_\alpha \Big[ a^\dagger_\alpha(\p)\dual{\lambda}^S_\alpha(\p) \exp( i p_\mu x^\mu)\nonumber \\
&& \hspace{11pt} + b_\alpha(\p)\dual{\lambda}^A_\alpha(\p) \exp(-i p_\mu x^\mu){\Big]}
\label{eq:elkob}
\end{eqnarray}
\end{subequations}
where $m$ is the mass of the ELKO field.
The creation and annihilation operators satisfy the following commutation relations
\begin{subequations}
\begin{eqnarray}
&& \left\{a_\alpha(\p),a^\dagger_{\alpha^\prime}(\p^\prime)\right\} = \left(2 \pi \right)^3 \delta^3\hspace{-2pt}\left(\p-\p^\prime\right) \delta_{\alpha\alpha^\prime}\label{eq:comm1} \\
&& \left\{a_\alpha(\p),a_{\alpha^\prime}(\p^\prime)\right\} = 0,\quad \left\{a^\dagger_\alpha(\p),a^\dagger_{\alpha^\prime}(\p^\prime)\right\} =0 \label{eq:comm2}
\end{eqnarray}
\end{subequations}
with similar relations for $b$'s. The spinors, $\lambda^S_\alpha$ and
$\lambda^A_\alpha$ are eigenstates of the charge conjugation operator, $C$,
such that
\begin{equation}
C\lambda^S_\alpha = +\lambda^S_\alpha\ \ \ \ \ \  C\lambda^A_\alpha = -\lambda^A_\alpha
\label{eq:conju}
\end{equation}
Here $\alpha$ is the helicity index.
The dual spinors are defined as, for example, 
\begin{eqnarray}
\dual{\mathfrak{\lambda}}^S_+(p^\mu) &=& -i\left[{\mathfrak{\lambda}}^S_-
\right]^\dagger \eta \nonumber\\
\dual{\mathfrak{\lambda}}^S_-(p^\mu) &=& i\left[{\mathfrak{\lambda}}^S_+
\right]^\dagger \eta
\end{eqnarray}
with similar relationships for the remaining spinors. The matrix $\eta$ is
given by,
\begin{equation}
\eta = \left(
\begin{array}{cc}
0 & 1\\ 
1 & 0
\end{array}   
\right)
\end{equation}
The spinors satisfy the following spin sums, 
\begin{eqnarray}
\sum_\alpha\dual{\mathfrak{\lambda}}^S_\alpha{\mathfrak{\lambda}}^S_\alpha= m(\mathbb{G(\phi)+I})\nonumber \\
\sum_\alpha\dual{\mathfrak{\lambda}}^A_\alpha{\mathfrak{\lambda}}^A_\alpha=m(\mathbb{G(\phi)-I})
\end{eqnarray}
where 
\begin{equation}
\mathbb{G(\phi)}= \left(
\begin{array}{cccc}
0 & 0 & 0 & -ie^{-i\phi}\\ 
0 & 0 & ie^{i\phi} & 0\\
0 & -ie^{-i\phi} & 0& 0 \\
ie^{i\phi} & 0& 0 & 0
\end{array}   
\right)
\end{equation}

The Lagrangian density for the ELKO field can be written as, 
\begin{equation}
\mathcal{L} = \partial^\mu\dual{\mathfrak{f}}\,\partial_\mu {\mathfrak{f}}(x) - m^2 \dual{\mathfrak{f}}(x) \mathfrak{f}(x) + \mathcal{L}_{\rm int}
\label{eq:L0}
\end{equation}
 where the interaction Lagrangian density is given by 
\citep{Ahluwalia:2013uxa} 
 \begin{eqnarray}
 \mathcal{L}_{int} &=&- g_{\mathfrak{f}\mathfrak{f}}(\dual{\mathfrak{f}}(x)\mathfrak{f}(x))^2 -g_{\mathfrak{f}\phi }\dual{\mathfrak{f}}(x)\mathfrak{f}(x) \phi^\dagger (x)\phi(x)\nonumber\\
& &- g_{\mathfrak{f}} \,\dual{\mathfrak{f}}(x) \left[\gamma_\mu,\gamma_\nu\right] \mathfrak{f}(x)F^{\mu\nu}(x)
\label{eq:Lint}
 \end{eqnarray}
and $g_{\mathfrak{f}\mathfrak{f}}$, $g_{\mathfrak{f}\phi}$ and $g_\mathfrak{f}$ are dimensionless coupling constants.
The first term on the right-hand side of Eq. \eqref{eq:Lint} represents the 
self-interaction of the ELKO field, the second is the interaction 
with the Higgs field, $\phi$ and the third its interaction with the electromagnetic
field \cite{Ahluwalia:2013uxa}. 

\section{ELKO as a cold dark matter candidate}
\label{sec:dark}
\begin{figure}[H]  
\begin{center}  
\includegraphics[width=0.30\textwidth]{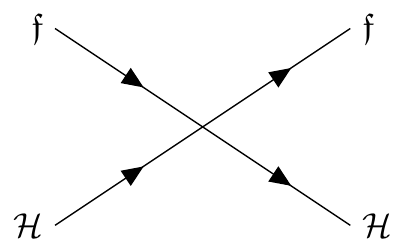}  
\caption{\small \sl ELKO-Higgs Scattering.  }
\label{fig:higgs}  
\end{center}
\end{figure}

\begin{figure}[H]  
\begin{center}  
\includegraphics[width=0.30\textwidth]{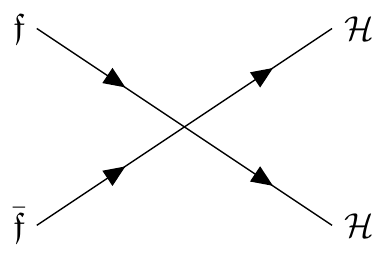}  
\caption{\small \sl Annihilation of ELKOs into a pair of Higgses.} 
\label{fig:higgs1}  
\end{center}
\end{figure}

If the ELKO-Higgs coupling,
$g_{\mathfrak{f}\phi}$, is significant then  
it could maintain these fermions in thermal equilibrium
with the cosmic plasma in the early Universe.  
The processes relevant for this purpose are shown in Figs. \ref{fig:higgs}
and \ref{fig:higgs1}. These correspond, respectively, 
to the ELKO-Higgs scattering
and the ELKO-ELKO annihilation into two Higgses. 

The amplitude of the ELKO-Higgs scattering process (Fig. \ref{fig:higgs}) is given by,
\begin{equation}
i\mathcal{M} =\frac{g_{\mathfrak{f}\phi}}{m}\dual{\lambda}^S_{\alpha^\prime}(k^\prime) \lambda^S_{\beta^\prime}(k) \,.
\end{equation}
This leads to,
\begin{equation}
|\mathcal{M}|^2 =\frac{g_{\mathfrak{f}\phi}^2}{m^2}4(EE^\prime-kk^\prime\cos(\theta-\theta^\prime))(1+\cos(\phi-\phi^\prime)) \,.
\end{equation}
The thermal averaged cross section for this scattering process is
\begin{equation}
\langle\sigma_s v\rangle= \frac{g_{\mathfrak{f}\phi}^2}{32\pi^2 m^2 s}\frac{1}{2} 4\pi(4EE^\prime-\pi kk^\prime \sin\theta) \,,
\end{equation}
where $E,E^\prime$ are initial and final energy of ELKO respectively. 
We assume an isotropic distribution of the ELKO momenta. Integrating over $\theta$, the thermally averaged cross section ($\sigma_s$)
 in the nonrelativistic limit is 
found to be,
\begin{equation}
\langle\sigma_s v\rangle = \frac{g_{\mathfrak{f}\phi}^2}{2\pi(m_H+m)^2}\,,
\label{E-higgs}
\end{equation}
where $m_H = 125$ GeV is the mass of the Higgs.

The amplitude of the pair annihilation process (Fig. \ref{fig:higgs1})
is given by
\begin{equation}
i\mathcal{M} =\frac{g_{\mathfrak{f}\phi}}{m}\dual{\lambda}^A_{\alpha^\prime}(k^\prime) \lambda^S_{\beta^\prime}(k) \,.
\end{equation}
The square of the invariant amplitude is
given by,
\begin{equation}
|\mathcal{M}|^2 =\frac{8g_{\mathfrak{f}\phi}^2}{m^2}(m^2+2{\bf{p}}^2)\,.
\end{equation}
In the nonrelativistic limit, this gives the following thermal averaged
annihilation cross section ($\sigma_a$), 
\begin{equation}
\langle\sigma_a v\rangle= \frac{g_{\mathfrak{f}\phi}^2}{16\pi m^2}\,.
\end{equation}

If ELKOs act as a CDM candidate, they will decouple from the cosmic
plasma when they are nonrelativistic. 
Let $T_f$ denote their freeze-out temperature \cite{Padma2010}.
Now, we use the fact that 
at the time of freeze-out, the interaction rate ($\Gamma$) 
becomes equal to the expansion rate ($H$), i.e. $\Gamma=H$.
Since, both the ELKO-Higgs scattering and the pair annihilation of ELKOs to
Higgses would contribute to the total thermally averaged
cross section at the time of decoupling of ELKO from
cosmic plasma,
the interaction rate is $\Gamma=n\langle\sigma v\rangle =
n(\langle\sigma_s v\rangle + n\langle\sigma_a v\rangle)$, where the 
number density $n$, 
in the nonrelativistic limit, is given by, 
$$n =g_A\Big(\frac{mT_f}{2\pi}\Big)^{\frac{3}{2}} \e^{-m/T_f}\,.$$
 Here $g_A$ is the degeneracy factor which is equal to 2 for ELKO. 
Now, the expansion rate or the Hubble constant can be expressed as, 
$$H(T_f) = 5.44 \frac{T^2_f}{M_{pl}}\,.$$ 
where $M_{pl}$ denotes the Planck mass.
Hence $\Gamma = H$ implies, 
\begin{eqnarray}
&&g_A\left(\frac{mT_f}{2\pi}\right)^{\frac{3}{2}}\exp[-m/T_f]\left(\frac{g_{\mathfrak{f}\phi}^2}{2\pi(m_H+m)^2}+\frac{g_{\mathfrak{f}\phi}^2}{16\pi m^2}\right)\nonumber \\
 &&\hspace{150pt} = 5.44\frac{T_f^2}{M_{pl}}
\label{H=T_higgs}
\end{eqnarray}

\begin{figure}[H]  
\begin{center}  
\includegraphics[width=0.50\textwidth]{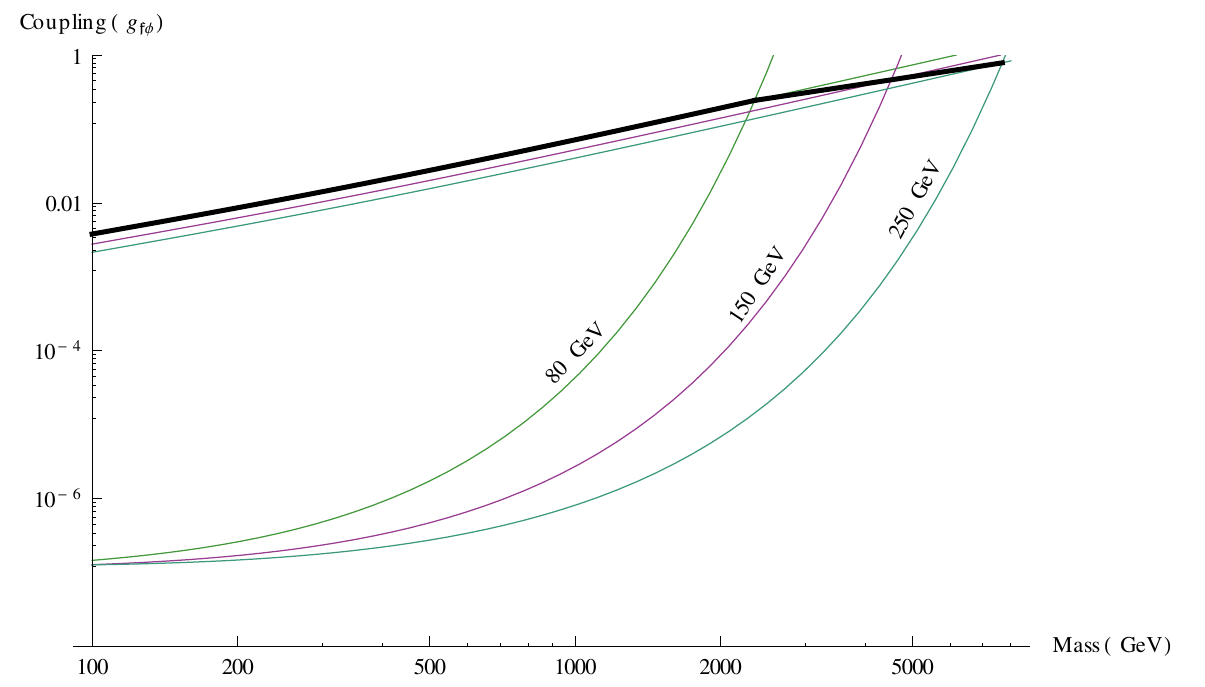}  
\caption{\small \sl The curved lines show the decoupling at different 
freeze-out temperatures, 250, 150 and 80 GeV. The lowest freeze out temperature
is 80 GeV since at this temperature the Higgs decouples from the plasma.  
 The slanted, almost straight lines are obtained by imposing the condition $\Omega_s =0.3$ for different freeze out temperatures. From bottom to top the 
decoupling temperatures are 250, 150 and 80 GeV. The dark line 
corresponds to the values of parameters for which ELKOs dominate the
dark matter density. The region above the dark line corresponds to the
 allowed range of parameters.  \label{graph:higgs}}.  
\end{center}
\end{figure}

In Fig. \ref{graph:higgs}, we plot the ELKO-Higgs 
 coupling $g_{\mathfrak{f}\phi}$ as a function of its mass, $m$ for a 
range of values of the decoupling temperature, $T_f$.
 We restrict the value of the coupling constant to be less than one so that
perturbation theory is applicable. 
The higher order corrections are suppressed by powers of $\alpha=g^2_{\mathfrak{f}\phi}/4\pi$ and hence are small, less than 10\%,
 as long as $g_{\mathfrak{f}\phi}<1$. 
We display the plots for $m\gtrsim 100$ GeV because the
 Higgs decouples from the cosmic plasma at a temperature of
around 80 GeV. 
 Hence, below this temperature
 ELKOs cannot maintain 
equilibrium with the cosmic plasma due to their interaction
with the Higgs. Only for mass much larger
than 100 GeV, ELKOs decouple as nonrelativistic particles and hence
act as CDM. 

We set the
 relic density of ELKO fermions equal to the dark matter density
$\Omega_s\approx0.3$, given by 
 \cite{Kolb:1989t},
\begin{equation}
\Omega_s=\frac{74.7 S_0 m}{2\pi^2 M_{pl}\sqrt{g_*}T_f \rho_c \langle\sigma v\rangle_f}
\label{omega}
\end{equation}
where $S_0 = 2.97\times10^3\ {\rm cm}^{-3}$ is the present value of entropy density, $\rho_c = 1.05\times 10^4h^2\ {\rm eV}/{\rm cm}^3$ is the critical density of the universe and we have assumed 
 $g_\ast = 106.75$, corresponding to the relativistic degrees of 
freedom at the time of decoupling. 
This leads to,
\begin{equation}
{m^3(m_H+m)^2\over    
(8m^2+(m_H+m)^2) T_f} =
3.37\times 10^8
g_{\mathfrak{f}\phi}^2 
\label{w=.3_higgs}
\end{equation}
in units of GeV$^2$.
This relationship between $m$ and $g_{\mathfrak{f}\phi}$ is also plotted in Fig.  \ref{graph:higgs} as slanted straight lines. 
For a given temperature, $T_f$, all parameter values below a particular
line will overclose the Universe and hence are ruled out.

The intersections of the two sets of curves 
 give the preferred range of the ELKO mass and its coupling with the Higgs. 
This is shown in Fig. \ref{graph:higgs} as a thick dark line. 
For these parameter values ELKO will give dominant contribution to
the cosmological dark matter density. 
For a given decoupling temperature, larger values of $g_{\mathfrak{f}\phi}$ are also
allowed but in this case we also require other dark matter particles
in order to fit the observed energy density.
Hence we find that ELKO acts as a CDM candidate if  
100 GeV $\lesssim m \lesssim$ 10,000 GeV and 0.005 
$\lesssim g_{\mathfrak{f}\phi} \lesssim$ 1.0.
As explained earlier, the upper limit on the coupling 
comes from the perturbative limit. 
Setting $g_{\mathfrak{f}\phi}=1$, we find that the corresponding value of ELKO
mass is approximately 10,000 GeV.

\begin{figure}[H]  
\begin{center}  
\includegraphics[width=0.30\textwidth]{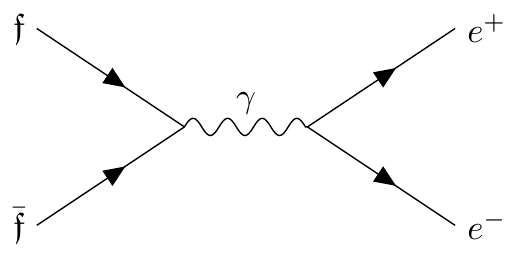}  
\caption{\small \sl Annihilation process for ELKOs.}
 \label{fig_electron}
\end{center}
\end{figure}

So far in this section, we have only considered the processes involving ELKO and the Higgs, but there is another type of process that, a $priori$, might also be relevant for
maintaining ELKO fermions in equilibrium with the cosmic plasma. An example of these
is shown in Fig. \ref{fig_electron}. These involve the coupling of ELKO with the electromagnetic field tensor. However,
as we shall see, direct dark matter searches impose severe restriction
on this coupling. Hence these processes do not give any significant contribution.

\section{Limits on ELKO from direct dark matter searches}
\label{sec:scatt}

We next consider the limits on ELKO couplings imposed by the direct  
dark matter searches using the CDMS II \cite{Agnese:2013rvf} results. 
For this purpose, 
we consider the scattering of ELKO with proton in nonrelativistic limit. 
We first determine the constraint on the ELKO-photon coupling and next
on the ELKO-Higgs coupling.

\subsection{Constraints on the ELKO-photon coupling}
\label{sec:const1}
\begin{figure}[H]
\begin{center}  
\includegraphics[width=0.25\textwidth]{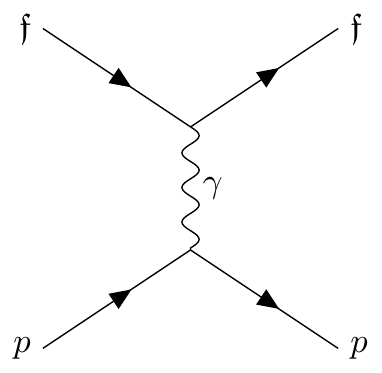}  
\caption{\small \sl ELKO-Proton Scattering by exchange of a photon.  }  
\label{fig:proton}
\end{center}
\end{figure}

\begin{figure}[H]  
\begin{center}  
\includegraphics[width=0.50\textwidth]{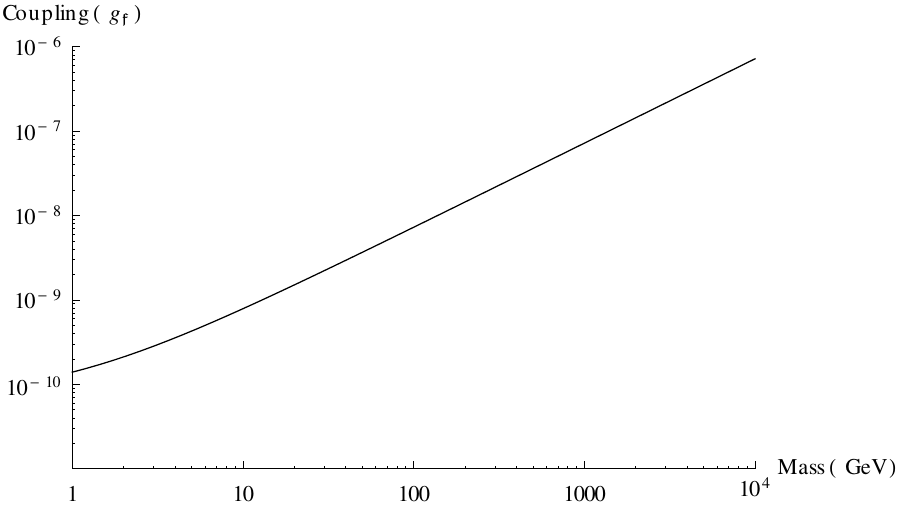}  
\caption{\small \sl The constraint imposed by
CDMS II on the ELKO electromagnetic coupling $g_\mathfrak{f}$ as a function  
of the ELKO mass.
 Only the region below the line is allowed.\label{graph:proton}}.  
\end{center}
\end{figure}

The dominant contribution to the ELKO-proton scattering due to the ELKO 
electromagnetic coupling is given by the 
 t-channel process, $\mathfrak{f}(k)p(p) \longrightarrow \mathfrak{f}(k^\prime)p(p^\prime)$, shown in Fig. \ref{fig:proton}.
The invariant amplitude for this process is given by
\begin{eqnarray}
i\mathcal{M} = && 4ig_{\mathfrak{f}}(\dual{\lambda}^S_{\alpha^\prime}(k^\prime) \sigma^{\mu\nu}\lambda^S_{\beta^\prime}(k))\frac{q_\mu g_{\nu\sigma}}{mq^2}\times\nonumber \\&&ie{\bar{u}}_{s^\prime}(p^\prime)[F_1\gamma^\sigma+\frac{\kappa}{2m_p} F_{2}i\sigma^{\sigma\alpha}q_\alpha]u_s(p) 
\end{eqnarray}
where $ F_1(q^2)$, $F_2(q^2)$ are the proton form factors, $m_p$ is the mass
of proton and $\kappa$ is the anomalous magnetic moment of proton. 
The momentum transfer in the process is $q = p^\prime-p$. The amplitude squared becomes
\begin{eqnarray}
|\mathcal{M}|^2 = &&\frac{16g_{\mathfrak{f}}^2e^2q_\mu q_\kappa}{m^2q^4}(\dual{\lambda}^S_{\alpha^\prime} \sigma^{\mu\nu}\lambda^{S}_{\beta^\prime})(\lambda^{S\dagger}_{\beta^\prime}\sigma^{\kappa\tau\dagger}\dual{\lambda}^{S\dagger}_{\alpha^\prime})\nonumber\\
&& \times Tr[(\slashed{p}^\prime +m_p)(F_1\gamma_\nu+\frac{\kappa}{2m_p} F_2i\sigma_{\nu}^ {\alpha}q_\alpha)\nonumber\\
&&\times(\slashed{p}+m_p)(F_1\gamma_\tau- \frac{\kappa}{2m_p} F_2i\sigma_{\tau}^{\rho}q_\rho)]
\label{eq:Msqr_proton}
\end{eqnarray}
Since ELKOs are dark matter candidates, we assume that they are moving
in random directions with respect to the Milky Way center. We consider an 
incoming proton, coming from the z-direction i.e. $p_{\mu}=(E_p,0,0,-p_3)$, 
with velocity $v=232$ km/s, which is equal to the speed of Sun around the
galactic center. We consider its scattering with an ELKO at rest.   
 The proton recoil energy turns out to be of order 10 KeV. 
 In the nonrelativistic limit, $F_1(q^2\approx 0)=1,F_2(q^2\approx 0)=1$. 
The scattering cross section in this limit is found to be, 
\begin{equation}
\sigma =\frac{\left(1.507\times 10^6+97382 \cos (\phi -\phi^\prime)\right)g_{\mathfrak{f}}^2}{m_p^2+m^2+2mE_p}
\label{eq:sigma_protn}
\end{equation}
where $\phi$ and $\phi'$ are the azimuthal angles of the momenta of the
initial and final
state ELKOs. We point out that for the initial state ELKO which is at rest,
we first assume a nonzero momentum that is later 
 set to zero.  
Integrating over $\phi^\prime$, the cross section becomes
\begin{equation}
\sigma =\frac{9.47\times 10^6 g_\mathfrak{f}^2}{m_p^2+m^2+2mE_p}
\label{eq:sigma_protn1}
\end{equation}
By using their silicon detectors, CDMS II \cite{Agnese:2013rvf} imposed
an upper-bound on the WIMP-nucleon scattering cross section $\sigma$ at $ 1.9\times 10^{-41}$ cm$^2$ (0.019 fb). The limits on the
coupling, $g_\mathfrak{f}$, for different ELKO masses,
shown in Fig. \ref{graph:proton}, are obtained by using the CDMS II limit in Eq.~\eqref{eq:sigma_protn1}. Only the region
below the line is allowed. For this range of parameters, we find that
the coupling $g_\mathfrak{f}$ gives negligible contribution for cosmic 
evolution 
of ELKOs. In order for this coupling to give a significant contribution to
the scattering cross section of ELKOs with cosmic plasma, its value would
have to be larger than 0.001 which is far above the limit allowed by CDMS II.
 Hence ELKO acts
as a dark matter candidate predominantly through its interaction with the Higgs.

\subsection{Constraints on the ELKO-Higgs coupling}
\label{const2}

\begin{figure}[H]
\begin{center}
\includegraphics[width=0.30\textwidth]{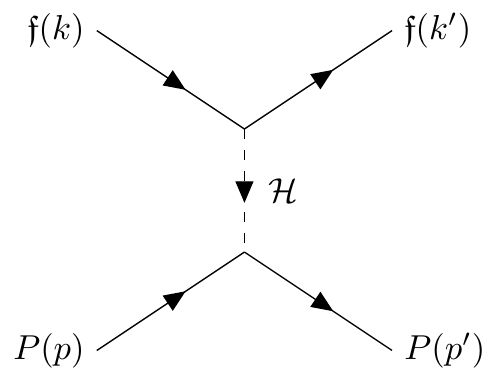}  
\caption{\small \sl Proton scattering with ELKO.\label{fig:p-elko-higgs}}
\end{center}
\end{figure}

\begin{figure}[H]
\begin{center}
\includegraphics[width=0.50\textwidth]{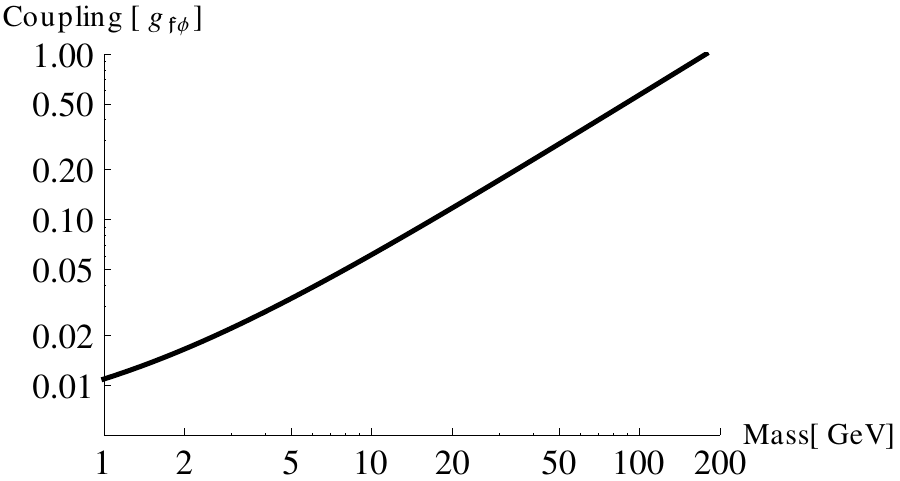}  
\caption{\small \sl The constraint imposed
by CDMS II on the ELKO-Higgs coupling, $g_{\mathfrak{f}\phi}$. 
For larger values of
the ELKO mass, the lower limit on $g_{\mathfrak{f}\phi}$ is larger than 1.  }
\label{fig:p-elko-higgs2}
\end{center}
\end{figure}

We next determine the constraint on the 
 ELKO-Higgs coupling, $g_{\mathfrak{f}\phi}$, imposed by CDMS II dark 
matter search.  
Expanding scalar field $\phi$ around the classical
ground state \cite{Alves:2014kta}
\begin{equation}
\phi = \frac{1}{\sqrt{2}}\left(H+v\right), \quad v = 246\ {\rm GeV},
\end{equation}
we obtain 
\begin{eqnarray}
\mathcal{L} &=& -\frac{1}{2}g_{\mathfrak{f}\phi}\dual{\mathfrak{f}}(x)\mathfrak{f}(x)H^2(x) - g_{\mathfrak{f}\phi} v \dual{\mathfrak{f}}(x)\mathfrak{f}(x)H \nonumber\\ &-&
\frac{1}{2}g_{\mathfrak{f}\phi}\dual{\mathfrak{f}}(x)\mathfrak{f}(x)v^2
\label{eq:Lag_Higgs}
\end{eqnarray}
The 2nd term in Eq. \eqref{eq:Lag_Higgs} gives the 3-point Higgs-ELKO-ELKO vertex. Using
this vertex we study the scattering of ELKO off proton in nonrelativistic
limits. The Feynman diagram for the ELKO-proton scattering with Higgs as intermediate particle is shown in Fig. \ref{fig:p-elko-higgs}. 
The amplitude for this process is given by
\begin{eqnarray}
i\mathcal{M} &=& \left(\frac{g_{\mathfrak{f}\phi} v}{m}\right)\left(\dual{\lambda}^S_\alpha(k^\prime)
\lambda^S_\beta(k)\right)\frac{i}{q^2 -m_H^2}\nonumber\\
& &\left(\frac{m_pF_H}{v}\right)
\bar{u}^{s^\prime}(p^\prime)u^s(p)
\label{eq:M_Higgs2}
\end{eqnarray}
where, as before, $q=p'-p$ is the momentum transferred.
The factor $m_pF_H/v$ is the low-energy 
effective coupling of the Higgs with proton. 
Here $F_H$ is the Higgs-proton form factor whose value has been estimated to be
approximately 0.35 \cite{Cline:2013,Mambrini:2011ik,Toussaint:2009pz,Young:2009zb,Ellis:2000ds} 
 in the limit $q^2\approx 0$. 
In the approximation $q^2 \ll m_H^2$, we have
\begin{eqnarray}
|\mathcal{M}|^2 &=& \frac{g_{f\phi}^2 m_p^2 F_H^2}{m_H^4 m^2}\Big
|\left(\dual{\lambda}^S_\alpha(k^\prime)
\lambda^S_\beta(k)\right)\left(\bar{u}^{s^\prime}(p^\prime)u^s(p)\right)\Big|^2\nonumber\\
\nonumber 
&=& \frac{g_{f\phi}^2 m_p^2 F_H^2}{m_H^4 m^2} 4\left(EE^\prime -kk^\prime
\cos(\theta-\theta^\prime)\right)\\
&& \times\left(1+\cos(\phi -\phi^\prime)\right)\times4(p.p^\prime +m_p^2) .
\label{eq:Msqr_Higgs2}
\end{eqnarray}
In the nonrelativistic limit the cross-section is given by
\begin{eqnarray}
\sigma &=& \frac{g_{\mathfrak{f}\phi}^2 m_p^2 F_H^2}{(64\pi^2s)(4m_H^4 m^2)}\nonumber\\
&&\times\ \Big[16\pi^2\left((4EE^\prime
-\pi kk^\prime \sin\theta)(8m_p^2)\right)\Big]
\end{eqnarray}
Assuming an isotropic incident ELKO flux, we obtain, after 
integrating over $\theta$, in the limit $k,k^\prime \rightarrow 0$, 
\begin{equation}
\sigma = \frac{4g_{\mathfrak{f}\phi}^2 m_p^4 F_H^2}{m_H^4(m_p+m)^2}
\label{eq:Cross_Higgs_2}
\end{equation}
Using the CDMS II constraint on this cross section, the limit imposed
on the coupling, $g_{\mathfrak{f}\phi}$, as a function of ELKO mass is shown in 
 Fig. \ref{fig:p-elko-higgs2}. The region below the line is the 
allowed range for the ELKO mass and the coupling, $g_{\mathfrak{f}\phi}$. 
We point
out that here also we have imposed an upper limit on the coupling such that,
 $g_{\mathfrak{f}\phi}<1$. The lower limit on  $g_{\mathfrak{f}\phi}$ turns
out to be greater than unity for larger values of the ELKO mass. 
As discussed earlier, with this constraint the higher order effects
are expected to be smaller than 10\%. 
We find that the CDMS II result does not produce any
 constraint on the parameter range, shown in Fig. \ref{graph:higgs}, for which
ELKO acts as a cold dark matter candidate. 

\section{Constraint on the ELKO$-$Higgs coupling from Gamma-Ray Bursts}
\label{sec:GRB}

The ELKO fermions break Lorentz invariance
due to the existence of a preferred axis. 
Hence they may induce Lorentz violating corrections in the photon
dispersion relation through loop effects. There exist stringent 
constraints \cite{Vasileiou:2013vra} 
on such effects due to data from gamma-ray bursts (GRBs) observed 
by Fermi$-$LAT \cite{0004-637X-697-2-1071}. In this section we determine
the constraints imposed by this data on the  
coupling of ELKO fermions with the Higgs particle. 
 
The modified photon dispersion relation in vacuum in a 
 Lorentz violating theory can be parametrized as 
\citep{Jacob:2008bw,Vasileiou:2013vra}
\begin{equation}
E^2=|\vec p\,|^2c^2\left[1-s_{\pm}\beta E^n\right]
\label{eq:lv1}
\end{equation}
where $E$ is the energy, $\vec p$ the three momentum, $\beta$ and $n$ 
parametrize the Lorentz violating effects and $s_\pm=\pm 1$ is the sign of 
Lorentz violation. If the Lorentz violation is attributed to quantum gravity
effects, then we have 
 $\beta=\frac{1}{(E_{QG})^n}$, where $E_{QG}$ is the scale of quantum
gravity. In the present case, however, $\beta$ is just a parameter
which characterizes the Lorentz violation contribution due to
ELKOs and has no apparent relationship to the scale of quantum gravity. 
The relationship in Eq. \eqref{eq:lv1},
implies that the photon group velocity depends upon the photon energy. 
Here the sign $s_\pm=-1(+1)$ corresponds to an increase (decrease) in photon velocity with an increasing photon energy. Hence, two photons of different energies, $E_h$ and $E_l$ ($E_h>E_l$) emitted by a distant point
source at the same instant will reach Earth with a time difference 
$\Delta t$. This time difference is related to the Lorentz invariance 
violation parameter $\tau_n$, defined as \citep{Jacob:2008bw},
\begin{equation}
\tau_n \equiv \frac{\Delta t}{E_h^n-E_l^n}\approx s_{\pm}\frac{\beta(1+n)}{2H_0}\times \int_{0}^{z} \frac{(1+x)^n dx}{\sqrt{\Omega_{\Lambda}+\Omega_M(1+x)^3}}\,.
\label{eq:delay}
\end{equation}
Here $H_0$ is Hubble constant, $z$ is redshift, $\Omega_M$ and $\Omega_{\Lambda}$ are matter and energy density respectively.
As explained in \cite{Vasileiou:2013vra} the data analysis can also be 
generalized to the case of a real GRB 
in which the photons are not emitted at the same time.

\begin{figure}
\begin{center}
\includegraphics[width=0.40\textwidth]{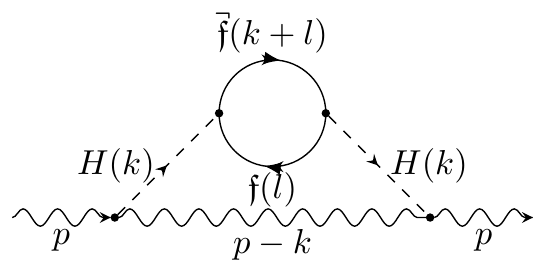}  
\caption{\small \sl Loop correction to the photon propagator}
\label{fig:loop1}
\end{center}
\end{figure}

 We next determine the change in the photon propagator due to an exchange
of the ELKO particles. The leading order diagram which contributes 
due to the Higgs-ELKO coupling is shown in Fig. \ref{fig:loop1}. 
The correction to the propagator leads to a modified dispersion relationship
for the photons, from which we can extract the parameters 
 $\beta$ and $n$ of Eq. \eqref{eq:lv1}. These can be used to
 calculate $\tau_n$ using the relation Eq. \eqref{eq:delay} for
the redshifts corresponding to different GRBs.

In the Higgs effective field theory$(heft)$ \cite{Alwall:2007st,Kniehl:1995tn,Shifman:1979eb}, the coupling of the Higgs with photons is mediated by 
top quark and W boson loops. The effective loop induced interaction 
Lagrangian can be written as, 
\begin{equation}
L_{heft}=-\frac{1}{4}gF_{\mu\nu} F^{\mu\nu} H\,,
\end{equation}
 where the coupling constant g is given by
 \begin{eqnarray}
 g=-\frac{\alpha}{\pi v}\frac{47}{18}\Big( 1+\frac{66}{235}\tau_w +\frac{228 }{1645}\tau_w^2+\frac{696}{8225}\tau_w^3\\ \nonumber 
 && \hspace{-178pt} +\frac{5248}{90475} \tau_w^4+\frac{1280}{29939}\tau_w^5+\frac{54528}{1646645}\tau_w^6-\frac{56}{705} \tau_t-\frac{32}{987}\tau_t^2\Big)
\end{eqnarray}                       
 Here $\tau_t=\frac{m_h^2}{4m_t^2}$ and $\tau_w=\frac{m_h^2}{4m_W^2}$. 
Hence the amplitude for the diagram shown in Fig. \ref{fig:loop1} can be
written as 
\begin{eqnarray}\label{eq:loop2}
\hspace{-5pt} i\Pi^{\mu\nu}= \frac{g^2g_{\mathfrak{f}\phi}^2 v^2}{4}\int\frac{d^4k}{(2\pi)^4}\frac{d^4l}{(2\pi)^4}\frac{I^{\mu\nu}}{(k^2-m_h^2+i\epsilon)^2}\\ \nonumber
&& \hspace{-170pt} \times \frac{Tr((\mathbb{G(\phi)+I})(\mathbb{G(\phi^{\prime\prime})+I}))}{(l^2-m^2+i\epsilon)((k+l)^2-m^2+i\epsilon)((p-k)^2+i\epsilon)}\,.
\end{eqnarray}
Here  $p^\prime=p-k$, $\phi$ and $\phi^{\prime\prime}$ are the azimuthal angles of $\vec{l}$ and $\vec{k}+\vec{l}$ respectively and 
\begin{eqnarray}
     I^{\mu\nu} &=& (p.p^\prime)^2 \left( g^{\mu\nu}-\frac{p^{\prime\mu}p^{\prime\nu}}{p^{\prime 2}}\right) 
 -(p.p^\prime)\left( g^{\nu\lambda}-\frac{p^{\prime\nu}p^{\prime\lambda}}{p^{\prime 2}}\right)(p^{\prime\mu}p_{\lambda})\nonumber \\
 &-&(p_{\sigma}p^{\prime\nu})\left( g^{\sigma\mu}-\frac{p^{\prime\sigma}p^{\prime\mu}}{p^{\prime 2}}\right)(p^{\prime}.p) \nonumber\\ 
&+&(p_{\sigma}p^{\prime\nu})\left( g^{\sigma\lambda}-\frac{p^{\prime\sigma}p^{\prime\lambda}}{p^{\prime 2}}\right)
(p^{\prime\mu}p_{\lambda}) \,.
\end{eqnarray}
The vacuum polarization $\Pi^{\mu\nu}$ satisfies the Ward identity, i.e. $p_{\mu}\Pi^{\mu\nu}(p)=0$. By gauge invariance, $\Pi^{\mu\nu}(p)$ is proportional to the $\left( g^{\mu\nu}-\frac{p^{\mu}p^{\nu}}{p^{2}}\right)$, i.e.
\begin{eqnarray}
\Pi^{\mu\nu}(p)=\left( g^{\mu\nu}-\frac{p^{\mu}p^{\nu}}{p^{2}}\right)\Pi(p^2,E).
\label{eq:Ward}
\end{eqnarray}
Using Eqs. \eqref{eq:loop2} and  \eqref{eq:Ward} , we find that the leading 
order correction to the propagator is given by, 
\begin{eqnarray}
 \hspace{-2pt}\Pi(p^2,E) \nonumber \\
&& \hspace{-10pt} =\frac{g^2g_{\mathfrak{f}\phi}^2 v^2}{6}\int\frac{d^4k}{(2\pi)^4}\frac{d^4l}{(2\pi)^4}\frac{(p.p^\prime)^2}{(k^2-m_h^2+i\epsilon)^2} \nonumber \\
&& \hspace{-10pt} \times \frac{Tr((\mathbb{G(\phi)+I})(\mathbb{G(\phi^{\prime\prime})+I}))}{(l^2-m^2+i\epsilon)((k+l)^2-m^2+i\epsilon)((p-k)^2+i\epsilon)}. \nonumber \\
\end{eqnarray}
It is not practical to use the standard Feynman parametrization for evaluating this 
integral because of the azimuthal angle dependent factors $\mathbb{G}(\phi)$ 
and $\mathbb{G}(\phi^{\prime\prime})$ in the numerator.
Instead, we use a different approach to estimate it. We     
     evaluate the $dl_0$ and $dk_0$ integral analytically
 and then, for different photon energies, integrate over $d^3\vec{k}$, $d^3\vec{l}$ numerically 
 using Monte-Carlo integration routine.

The correction to the dispersion relation, $-s_\pm\beta E^n$ is 
equal to $\Pi(p^2,E)/|\vec p\,|^2$. 
Since this term is expected to be a small correction, we can consistently
set $p^2 = E^2-|\vec p\,|^2=0$ in its evaluation.
We are primarily interested in the energy regime 0.1 GeV $<E<$ 10 GeV
which overlaps closely with the range of energy of the events observed
in GRBs 080916C, 090510, 090902B, 090926A
studied in \cite{Vasileiou:2013vra}. 
The result depends on the direction of photon propagation since the
basic framework violates Lorentz invariance through the appearance of
factors, such as, $\mathbb{G}(\phi)$, in the spin sums. 
However we find that the result does not depend qualitatively on the
direction of propagation and fix the direction such that in our chosen
frame the spherical polar coordinates of the photon momentum are
$\theta=\phi=\pi/4$. We have verified that the order of magnitude of the
final answer does not change with choice of propagation direction.
We find that for $E\ll 0.1$ GeV
and for $E\gg 10$ GeV, the correction factor is almost independent of
energy, i.e. corresponds to $n=0$. However in the range  0.1 GeV $<E<$ 10 GeV
we find a small decrease in the correction factor. 
We restrict ourselves to this energy range while determining
the effective value of $n$.  
We define a parameter $\beta'$ such
that $\beta=\beta'g_{\mathfrak{f}\phi}^2$.   
The resulting 
extracted values of $\beta'$ and $n$ for different choices of ELKO masses
are given in Table \ref{tab:parametersPi}. 
  
\begin{table}[h]
 \begin{tabular}
{ |p{2.5cm}|p{3cm}|p{2.5cm}| }
 \hline
 Mass(m) in GeV   & $\beta'$ & $n$ \\
 \hline
 100  & $1.33\times 10^{-6}$    &-0.18\\
 500& $1.30\times 10^{-6} $ &-0.12 \\
 1000 & $1.22\times 10^{-6}$ & -0.12 \\
 2000   & $9.41\times 10^{-7}$ & -0.12\\
 5000 &    $7.05\times 10^{-7}$  & -0.12\\
 9500 &  $5.39\times 10^{-7}$  & -0.12  \\
 \hline
\end{tabular}
\caption{Parameter $\beta'$ and $n$ for different ELKO masses.}
\label{tab:parametersPi}
\end{table} 

The fact that our Lorentz violating correction to the dispersion relations
is not proportional to either $E$ or $E^2$, as is often assumed 
within the framework of quantum gravity
\cite{Vasileiou:2013vra,Jain:2005}, is not surprising. The current framework is
closest to the Very Special Relativity (VSR) invariant theories which tend to show dominant
deviation from Lorentz violation at low energies due to the 
presence of nonlocal contributions \cite{Cohen:2006ky,Nayak:2015}. The ELKO
framework is somewhat unique since the Lorentz violating terms appear
explicitly only in the spin sum and not the action. Hence it is expected
to deviate both from the VSR invariant theories, as proposed in
\cite{Cohen:2006ky}, and the expectation that Lorentz violating
effects might increase with energy as $E$ or $E^2$ due to quantum gravity
effects. 

We next use the GRB data to impose a limit on the parameter
$\beta$ and hence on the ELKO coupling,
$g_{\mathfrak{f}\phi}$. A detailed data analysis for this purpose is 
rather complicated and beyond the scope of the present paper.  
Here we restrict ourselves to extracting an order of magnitude
estimate of the limit. 
In Ref. \cite{Vasileiou:2013vra}, GRB data was 
used in order to impose a limit on the quantum gravity scale $E_{QG}$ for
$n=1,2$. They used the GRBs  
080916C, 090510, 090902B, 090926A for this calculation. 
The data for
these bursts is mostly confined to energies less than 10 GeV. 
In fact most of the data lies in the range $E<1$ GeV and in  Ref. \cite{Vasileiou:2013vra} the authors impose a lower limit of 30 MeV. 
Here we directly use their extracted value of $\tau_n$ with $n=1$ and make an
estimate of $\Delta t$ setting $E_h=1$ GeV and $E_l=0.1 $ GeV. 
For all the GRBs it is found that $|\tau_n|\lesssim 1$ s/GeV.  
Hence we set $|\tau_n|\approx 1$ s/GeV which leads to $\Delta t\approx 1$ sec. 
We find that the 
extracted value of $\Delta t$ does not show a strong dependence on
 $E_l$ or the
chosen value of $n$, i.e. the value obtained with $n=2$ is not too 
different from that corresponding to $n=1$. Using
this value of $\Delta t$ in Eq. \ref{eq:delay}, and the $\beta'$ and $n$ values
given in Table  \ref{tab:parametersPi} we obtain an order of magnitude
estimate of the limit on $g_{\mathfrak{f}\phi}$. 

We find that for all the GRBs, 
080916C, 090510, 090902B, 090926A and for the entire range of ELKO mass values
given in Table  \ref{tab:parametersPi}, the limiting values of 
$g_{f\phi}$ lie in the range $10^{-5}$ to $10^{-6}$. Hence we obtain 
a conservative upper limit $g_{\mathfrak{f}\phi}<10^{-5}$. 
This implies that the
 Fermi-LAT data actually rules out the parameter space we obtained 
by demanding that ELKO acts as a CDM candidate subject to the limits imposed 
by direct detection experiments. Hence we conclude that 
ELKO fermion cannot be a 
dominant CDM candidate. It can of course still contribute
as a subdominant cold dark matter candidate. Alternatively ELKOs might
never have been in equilibrium with the cosmic plasma. In this case they
may still contribute significantly to the energy density of the dark matter despite
the limit due to GRBs.
However we have not investigated this in this paper. 


\section{Limit on the self-coupling from astrophysical data}
\label{sec:self-interaction}
In the earlier sections we have shown that cosmological and astrophysical
observations lead to severe constraints on the coupling of ELKO fermions
with Higgs and photons. Indeed the entire parameter regime for which it
can contribute significantly as a cold dark matter candidate is ruled out. 
As already mentioned, the only allowed
possibility is
that ELKO fermions were never in equilibrium with the cosmic plasma.
In this case the constraints imposed by cosmological considerations
(see Fig. \ref{graph:higgs}) are not applicable and such fermions
can contribute significantly to nonrelativistic dark matter density. 
In this section
we assume such a scenario and determine implications of the ELKO  
self-interaction term of Eq. \eqref{eq:Lint}, 
$g_{\mathfrak{f}\mathfrak{f}}(\dual{\mathfrak{f}}(x)\mathfrak{f}(x))^2$, for the nonrelativistic 
dark matter cores in galactic halos. 

 The self-interacting nonrelativistic dark matter has 
been proposed to solve the problem with the small scale structure formation 
of the Universe.  
The density of dark matter cores in the galactic centers is observed to
be lower than the value predicted by weakly interacting
nonrelativistic dark matter. The lower density can be explained by invoking
 collisional (self-interacting) dark matter. In this model, the dark matter has large scattering cross section and negligible annihilation rate. Assuming that the ELKO particles are nonrelativistic, the scattering cross section $\dual{\mathfrak{f}}\mathfrak{f} \rightarrow \dual{\mathfrak{f}}\mathfrak{f}$ is 
\begin{equation}
\sigma_{\dual{\mathfrak{f}}\mathfrak{f}} = \frac{g_{\mathfrak{f}\mathfrak{f}}^2}{4\pi m^2}\,.
\label{eq:self-int}
\end{equation}

For this scenario to work, the mean free path $(\lambda)$ of the collisional 
dark matter should be in the range of 1 Kpc to 1 Mpc at the location of the
Sun within the Milky way. Here the   
mean density of dark matter is 0.4 GeV/cm$^3$ 
\citep{Spergel:1999mh,Bento:2001yk}. 
Using the result for the elastic scattering cross section for such 
a dark matter \citep{Spergel:1999mh} 
and applying this for the ELKO-ELKO scattering 
we obtain 
\begin{equation}
\sigma_{\dual{\mathfrak{f}}\mathfrak{f}}= 8.1\times 10^{-25} cm^2 \left(\frac{m}{GeV}\right) \left(\frac{\lambda}{1 Mpc}\right)^{-1}\,.
\label{eq:elastic}
\end{equation}
From Eqs. \eqref{eq:self-int} and \eqref{eq:elastic}, we get  
\begin{equation}
g_{\mathfrak{f}\mathfrak{f}}=161.71\times  \left(\frac{m}{GeV}\right)^{3/2} \left(\frac{\lambda}{1 Mpc}\right)^{-1/2}\,.
\end{equation}
   Typical range of self-interacting dark matter mass is 1 MeV to 10 GeV \citep{Spergel:1999mh}, so depending upon the mean free path, the ELKO self-coupling is constrained by the above relation.
In particular as we vary $\lambda$ from 1 Mpc to 1 Kpc the minimum
value of coupling $g_{\mathfrak{f}\mathfrak{f}}$ is found to vary
from 0.005 to 0.16. These values are obtained by setting the ELKO mass
$m=1$ MeV. For the range of $\lambda$ and $m$ values quoted above, the
upper limit on the coupling exceeds unity.

\section{Conclusion}
\label{sec:conclusions}
The ELKO fermion is an interesting and natural dark matter candidate. 
By its very existence it violates Lorentz invariance
and respects only a subgroup.  
By its intrinsic nature, its interactions with most of the 
standard model fields 
are severely restricted. 
It couples dominantly with the Higgs particle. Hence, 
in the ELKO proposal we find an interesting
prediction that the dark matter sector as well as its coupling to Higgs
must violate Lorentz invariance. In the present paper we have made a 
detailed analysis of the implications of ELKO fermions as a cold dark matter 
candidate. 
We find that ELKO acts as a cold dark matter candidate 
if its mass lies in the range 100 to 10,000 GeV. 
The upper bound on 
ELKO mass is obtained by demanding that the  
 Higgs-ELKO coupling $g_{\mathfrak{f} \phi}<1$, that is, it
stays within the perturbative regime. 
Below the lower limit  
 it will not decouple
from cosmic plasma as a nonrelativistic particle.  
   The lower limit on the coupling $g_{\mathfrak{f} \phi}$
is found to be 0.005.
However this entire range of coupling is eliminated by the constraint
imposed by time delay observations of photons of different 
energies emitted in gamma ray bursts. 
This constraint arises since the ELKO fermion induces a Lorentz violating
term in the photon dispersion relations. Such a term leads to a delay
in arrival times of photons of different energies emitted by 
gamma ray bursts and hence is constrained by the observed
time delay. 
Hence we conclude that ELKO does not contribute significantly as a 
cold dark matter candidate. However it may still contribute significantly
to dark matter if it were never in equilibrium with the cosmic plasma.

ELKOs also couple to photon via nonstandard $g_{\mathfrak{f}} \,\dual{\mathfrak{f}}(x) \left[\gamma_\mu,\gamma_\nu\right] \mathfrak{f}(x)F^{\mu\nu}(x)$ interaction. We find that this coupling is severely constrained by direct dark matter search
experiments, such as CDMS II. However we find that CDMS II does not impose
a significant constraint on the ELKO-Higgs coupling.  

Finally we have obtained the range of values for the ELKO mass and self- 
coupling for which it may be consistent with the density of
dark matter core in the galactic center. This requires the dark matter
to have significant cross section for scattering with other dark matter
particles. Hence it provides us with a handle on the self-coupling. 

\section{Acknowledgment}
We thank Gopal Kasyap for useful discussions.
\bibliographystyle{apsrev}
\bibliography{Reference}

\end{document}